\documentclass[final,3p,times,onecolumn]{elsarticle}

\usepackage{graphicx}
\usepackage{epsfig}
\usepackage{amsmath,amsfonts, amsthm, mathrsfs, amssymb, bbm}
\usepackage{mathtools}
\usepackage{commath}
\usepackage[sc,osf]{mathpazo}
\usepackage[ruled,vlined,noresetcount]{algorithm2e}
\usepackage{natbib}
\usepackage{times}
\usepackage{subfigure}
\usepackage{lineno}
\graphicspath{ {images/} }
\usepackage{multirow}
\journal{Applied Mathematics and Computation}

\begin{document}

\begin{frontmatter}

\title{Sentiment-Based Prediction of Alternative Cryptocurrency Price Fluctuations Using Gradient Boosting Tree Model}

\author[DMath]{Tianyu Ray Li}
\ead{tianyu.li.20@dartmouth.edu}
\author[DMath]{Anup S. Chamrajnagar\corref{ff}}
\ead{anup.s.chamrajnagar.18@dartmouth.edu}
\author[DMath]{Xander R. Fong}
\ead{xander.R.Fong.18@dartmouth.edu}
\author[DMath]{Nicholas R. Rizik}
\author[DMath,DMed]{Feng Fu\corref{ff}}
\ead{fufeng@gmail.com}

\address[DMath]{Department of Mathematics, Dartmouth College, Hanover, NH 03755, USA}

\address[DMed]{Department of Biomedical Data Science, Geisel School of Medicine at Dartmouth, Lebanon, NH 03756, USA}

\cortext[ff]{Corresponding author at: 27 N. Main Street, 6188 Kemeny Hall, Department of Mathematics, Dartmouth College, Hanover, NH 03755, USA. Tel.: +1 (603) 646 2293}

\begin{abstract}
In this paper, we analyze Twitter signals as a medium for user sentiment to predict the price fluctuations of a small-cap alternative cryptocurrency called \emph{ZClassic}. We extracted tweets on an hourly basis for a period of 3.5 weeks, classifying each tweet as positive, neutral, or negative. We then compiled these tweets into an hourly sentiment index, creating an unweighted and weighted index, with the latter giving larger weight to retweets. These two indices, alongside the raw summations of positive, negative, and neutral sentiment were juxtaposed to $\sim 400$ data points of hourly pricing data to train an Extreme Gradient Boosting Regression Tree Model. Price predictions produced from this model were compared to historical price data, with the resulting predictions having a 0.81 correlation with the testing data. Our model'€™s predictive data yielded statistical significance at the  $p < 0.0001$ level. Our model is the first academic proof of concept that social media platforms such as Twitter can serve as powerful social signals for predicting price movements in the highly speculative alternative cryptocurrency, or ``alt-coin'', market.

\end{abstract}


\end{frontmatter}


\section{Introduction}
\label{intro}

A cryptocurrency (or crypto currency) is a digital asset designed to work as a medium of exchange that uses cryptography to secure its transactions, control the creation of additional cryptocurrencies, and verify the secure transfer of assets~\cite{Greenberg}. Cryptocurrencies can be classified as types of digital or alternative currencies, distinct from traditional currencies in that they are founded on the principle of decentralized control, compared to the central banking systems that typical currencies rely on~\cite{Allison}. The inception of cryptocurrencies dates back to 2008, when an unknown entity under the pseudonym Satoshi Nakamoto publically released a paper titled Bitcoin: A Peer-to-Peer Electronic Cash System~\cite{Nakamoto}. In January 2009, Nakamoto implemented the bitcoin software as open source code, releasing it to the public on \emph{SourceForge}~\cite{Davis}. Nakamoto'€™s contributions galvanized a wave of public attention, spurring others to create alternative cryptocurrencies that relied on the same fundamental technology but were specialized in purpose~\cite{ElBahrawy_RSOS17}. 

This wave of new cryptocurrencies has received much attention by the media and investors alike due to the assets'€™ innovative features, potential capability as transactional tools, and tremendous price fluctuations. In the past two years, the total market capitalization of the entire cryptocurrency market has increased 11,600\% from $\$7.4$ billion dollars in January 2016 to over $\$800$ billion dollars as of January 2018~\cite{zcl}. This exponential growth is the result of both increased investor speculation and the introduction of various new cryptocurrencies, with current estimates of the total number of cryptocurrencies topping $1,400$ different coins~\cite{Cornish}. Thus, analyzing evolutionary dynamics of the cryptocurrency market is a topic of current interest and can provide useful insight about the market share of cryptocurrencies~\cite{ElBahrawy_RSOS17}. Moreover, longitudinal   datasets of Bitcoin transactions have been used to identify the socio-economic drivers in cryptocurrency adoption~\cite{Parino_arxiv}. 

The speculation behind these digital assets has increased to such magnitudes that even cryptocurrencies with no functionality have surpassed the market value of established companies whose stocks are publicly traded in the equity markets. This rapid and exponential increase in cryptocurrency prices suggests that price fluctuations are driven primarily by retail investor speculation, and that this market exhibiting signs of a financial bubble~\cite{Macedo}. In light of this, a recent study quantifies the inefficiency of the Bitcoin market by studying the long-range dependence of Bitcoin return and volatility from 2011 until 2017~\cite{Bariviera_EL17}. Noteworthy, there has been increasing attention paid to improving our understanding of cryptocurrency market behavior, for example, by means of field experiments of peer influence (exerted by bots) on human trading decisions~~\cite{Krafft_arxiv} and probabilistic modeling of buy and sell orders~\cite{Guo_arxiv}.

Given that the alternative cryptocurrency market is dominated by retail investors, with few large institutional investors, sentiment on social media platforms and online forums may present a viable medium to capture total investor sentiment~\cite{Leinz}. More recently, it has been shown that social media data such as Twitter can be used to track investor sentiment, and price changes in the Bitcoin market and other predominant cryptocurrencies~\cite{Meucci,Gacia_RSOS15,Kim_PLoS16,Phillips_CI17}. In Ref.~\cite{Gacia_RSOS15}, the authors demonstrate that Twitter sentiment, alongside economic signals of volume, price of exchange for USD, adoption of the Bitcoin technology, overall trading volume could be used to predict price fluctuations. 

As a consequence, investors may have adopted a similar strategy within the Bitcoin market, thereby weakening the correlation between Twitter sentiment and Bitcoin prices. Moreover, the daily trading volume of cryptocurrencies has increased such that conditions are now suitable for high-frequency trading firms to exploit this correlation~\cite{Williams-Grut}. For proof of concept, we thus decided that using Twitter sentiment to analyze price fluctuations of nascent alternative cryptocurrencies (commonly known as ``alt-coins''€) could provide valuable insight, and eventually lead to a viable arbitrage opportunity in other emerging alternative cryptocurrencies. Therefore, we aim to analyze and build a machine learning pricing model for this highly speculative market through gauging investor sentiment via Twitter, a pervasive social network that has been strongly suggested to serve as a powerful social signal for Bitcoin prices~\cite{Gacia_RSOS15}. 

\section{Material and methods}

We began by researching different alternative cryptocurrencies to ultimately decide which would be best suited within the confines of our analysis. Ultimately, we decided to choose ZClassic (ZCL), a private, decentralized, fast, open-source community driven virtual currency, as the primary target of our academic focus given its unique technological dynamics and suitability of trading volume within the confines of our computational capacity. First off, the technological nature of the ZClassic cryptocurrency lends itself to a high level of predictability via tweet analysis. Specifically, ZClassic is set to €œ``hard fork€'' into Bitcoin Private on February 28th, 2018. A €œhardfork is a major change to blockchain protocol which makes previously invalid blocks or transactions valid~\cite{Hayes}.

As a result, the single cryptocurrency (ZClassic) preceding the hard fork will be split into two, ZClassic and Bitcoin Private~\cite{Hayes}. Previous hardforks include Bitcoin Cash and Bitcoin Gold, and the history of each suggests that ZClassic'€™s price fluctuations will be largely based off speculation regarding the future success and accessibility of Bitcoin Private. For example, any news release that is seen by investors as indicative of the possibility that Bitcoin Private will be traded on a major exchange or that the fork will be supported by a certain exchange will exert upwards price pressure on the cryptocurrency'€™s price. As such, real-time tweet analysis serves as a suitable means to gauge investor sentiment following these news releases, and pinpoint spontaneous news releases themselves. Secondly, the relatively lower trading volume of ZCL compared to that of alternative cryptocurrencies suggests that it may be more susceptible to sentiment-based price movement. 

To collect the tweets, we decided to base our program in RStudio, given its motley of free Twitter-analysis packages and foundations within data analysis and statistical computing. Specifically, we used the open-sourced rtweet package~\cite{rtweet}, which accesses Twitter'€™s REST and stream APIs. We were able to use the rtweet package to retrieve, from each of the last seven days, searching from midnight backwards, tweets that had the terms ``ZClassic,''€ ``€œZCL,''€ and ``€œBTCP.''€ This collection process was repeated 3 times over the course of three and a half weeks to provide sufficient data for our analysis. We then merged all data sets, and eliminated any duplicate tweets given that a single tweet could contain all three of these terms and therefore be accounted for thrice in the final data set. In the end, we garnered a final data set of $130,000$ unique tweets. 

We then created an algorithm to classify each tweet as positive, negative, or neutral sentiment using natural language processing. The dictionary, primarily sourced from the Python package ``Textblob'', that assigns impactful words and phrases a polarity value (e.g., ``€œtop''€ and ``€œnot great''€ have values of $0.5$ and $-0.4$, respectively), which we view as sentiment. Thus, each tweet is assigned a polarity value between $-1$ and $1$ based on the combinations of keywords and phrases. If the entire tweet string has a positive nonzero polarity value, our program scores the sentiment as positive, or $+1$. If the entire tweet string has a negative nonzero polarity value, our program scores the sentiment as negative, $-1$. If the polarity value is zero, then the tweet receives a sentiment value of $0$.

Another important aspect to note regarding the character of each tweet is the chained network effect that each retweet creates. It is evident that retweets can cause a chain effect, thereby increasing the dispersion of the initial ``€œtweet.''€ As such, it is possible that retweeted posts contain new positive or negative information, or may be viewed by the trading community as ``€œinsightful.''€ For this reason, we decided to create a second sentiment index in which retweets would be more heavily weighted than tweets themselves, using it as one of the features in training our model. We respectively assigned a weight of $-2$ or $+2$ to every negative and positive retweet because we assumed retweets signify more newsworthy events and have greater credibility than single tweets. Thus, we believe cryptocurrency investors will be more likely to react to retweets than to single tweets. Both the values of our weighted and unweighted sentiment indices were then calculated on an hourly basis by summing the weights of all coinciding tweets, which allowed us to directly compare this index to available ZCL price data.

For model selection, we employed $10$-fold cross validation on $589$ data points to choose an optimal model framework among linear regression, logistic regression, polynomial regression, exponential regression, tree model, and support vector machine regression. A tree model called the Extreme Gradient Boosting Regression (also known as XGBoost~\cite{Chen_KDD}), exhibited the smallest loss, or inaccuracy, and was thus chosen to train the model on our data. The XGBoost model, as well as other tree-based models, is particularly suited for applications on our data for the following reasons:
\begin{enumerate}
    \item Tree models are not sensitive to the arithmetic range of the data and features. Thus, we do not need to normalize the data and possibly prevent loss due to normalization. 
    \item Tree models are by far the most scalable machine learning model due to their construction processes - simply adding more children nodes to the pre-existing tree nodes will update the tree and allow our strategy to continue to accurately predict price as our collection of price and tweet data increases into the future. It also makes the model adaptable for currencies with larger daily tweet volumes. 
    \item On the abstract level, the tree model is a rule-based learning method which, unlike a traditional regression learning method, has more potential to unveil insightful relationships between features.
\end{enumerate}
XGBoost is a tree ensemble model, which outputs a weighted sum of the predictions of multiple regression trees, by weighing mislabeled examples more heavily. 

For completeness, we sketch the key ideas behind XGBoost as follows. Let us define
$$\hat{y_i}=\phi(x_i) = \sum_{k=1}^{K} f_k(x_i),\quad f_k\in \mathcal{F}, $$
where $\hat{y_i}$ is the prediction from our model for the $i$-th observation, $\phi(x_i)$ is our predicting function
and each $f$ representing a tree in our regression tree forest, $\mathcal{F}$. Our goal is to minimize the objective function $\mathcal{L}$ , defined below:
$$\mathcal{L}(\phi) = \sum_{i}l(\hat{y_i},y_i)+\sum_{k}\Omega(f_k),$$
where $$\  \Omega(f) = \gamma\mathcal{T}+\frac{1}{2}\lambda \norm{w}.$$
The function $l(\hat{y_i},y_i)$ represents a loss function, which in this case is a mean-square function, and the $\Omega(f_k)$ is a regularization, which penalizes each tree for having too many leaves and to ensure
smooth final learned weights. The definition of this regularization follows the above equation where $w$ is the coefficient at each node and $\mathcal{T}$ is the number of leaves in the tree.

To minimize the above objective function, we employed a greedy Algorithm~\eqref{xgboost} to create our regression tree forest $\mathcal{F}$ as originally implemented in Ref.~\cite{Chen_KDD}.

\begin{algorithm}[h]
\SetAlgoLined
\KwIn{$I$, instance set of current node}
\KwIn{$d$, feature dimension}
$gain \leftarrow 0$\\
$G \leftarrow \sum_{i\in I} g_i$, $H \leftarrow \sum_{i\in I} h_i$\\
 \For{$k = 1$ to $m$}{
 $G_L \leftarrow 0$, $H_L \leftarrow 0$\\
 \For{$j$ in sorted $(I, \mbox{by}\, \mathbf{x}_{jk})$}
 {
 $G_L \leftarrow G_L + g_j, H_L \leftarrow H_L + h_j$\\
  $G_R \leftarrow G - G_L, H_R \leftarrow H - H_L$\\
 $  score \leftarrow  \max(score, \frac{G_L^2}{H_L+\lambda} + \frac{G_R^2}{H_R+\lambda} - \frac{G^2}{H+\lambda}$ )
 }
 }
 
 \KwOut{Split with max score}

 \caption{Exact greedy algorithm for split finding~\cite{Chen_KDD} used in our price prediction model.}\label{xgboost}
\end{algorithm}

One-third of the $589$ data points is separated as the testing data, and the remainder is used as the training set as we built our Extreme Gradient Boosting Regression model. The model also tests different lead-lag on the range of $[0,\,1,\,2,\,3,\,4,\,5\, \text{hours}]$ since we do no€™t know how quickly the public would react to the market update or the social media sentiment. Based on the testing result, we decided that there is a $3$-hour lag effect between social media information and price effects.

\section{Results}

To begin, our natural language processing classification algorithm showed significant accuracy in identifying the sentiment of each tweet (see Table 1). Examples of tweets that received positive, neutral, and negative sentiment values are shown in Table 2.
\\
\\ 
\newcommand\measureISpecification{4ex}
\newcommand{\ctab}[1]{\raisebox{\dimexpr \measureISpecification/2 -.748ex}{#1}}
\begin{table}
\centering
\begin{tabular}{rr|ccc}
   && \multicolumn{3}{c}{Algorithm Sentiment Prediction}\\
   && Positive & Neutral & Negative \\
    \hline 
   &&\\[-2ex]
   & \ctab{Positive} & 79\% & 21\% & 0\% \\
   \raisebox{\dimexpr \measureISpecification/2}[0pt][0pt]{{\small Manual Sentiment Decision}} & \ctab{Neutral} & 34\% & 51\% & 15\% \\
   & \ctab{Negative} & 25\% & 0\% & 75\%
\end{tabular}
\caption{Validation analysis of algorithm sentiment prediction by manual inspection.}\label{manualinsep}
\end{table}

\begin{table}[ht]
\begin{center}
\begin{tabular}{p{1.5cm}|p{12cm} }
\hline
Sentiment & Tweets \\\hline
\multirow{3}{2em}{Positive} &  ``Bitcoin Private \$btcp is coming! better hop in \$zcl to grab some before the snapshot on the 28th! 15 days is a times for a crypto to be going up.''  \\
 & ``\#ZCL going to give Grandma her retirement back!''\\
 & ``If you haven't got \$ZCL yet, now is the time to grab some.''\\\hline
\multirow{3}{2em}{Neutral} & ``Do you plan on selling your \$zcl before the fork or holding on to it for the \$btcp?''\\
& ``don't miss out registering on Binance, before they close registration again.''\\
& ``RT @cryptobr4in: When everyone realizes your get FREE \#BitcoinPrivate \$BTCP if you buy \$ZCL \#CryptoCurrency \#FreeCoins ...''\\\hline
\multirow{3}{2em}{Negative} & ``RT@Crypto\_Bitlord: Bitcoin private is just another scam coin knock off like bitcoin cash trying to ride the satoshi's original vision.''\\
& ``2017 end - most thought 2018 would b the year crypto would explode. I said it b4 n will say it again - gonna b a bad year.''\\
& ``What a joke this \$zcl game is. Good news, drops like a rock. Fucking whales. \#cryptocurrency ...''\\ \hline
\end{tabular}
\caption{Examples of Tweets with positive, neutral and negative sentiment classifications in our dataset.}
\end{center}
\label{tweetsenti2}
\end{table}

Upon reviewing our data set of tweets, one major concern we had was the flood of computer-generated €œ\emph{bot}€ tweets, which often promote contests and giveaways. In practice, retail investors often ignore these tweets, given their obvious usage as means of commercial promotions. These are often written using positive language; however, the vast majority of these were properly characterized as neutral. To further gauge the accuracy of our algorithm, we manually classified a sample of 100 random tweets, comparing them to our algorithm'€™s classifications to measure false classification rates. Table 1 shows the general distinctions between our algorithm'€™s classifications and manual classifications.

In all three cases we can see that the chance of the algorithm guessing the sentiment correctly is over $50\%$. The algorithm boasts a near $80\%$ success rate in successfully classifying positive tweets, and correctly characterized $0\%$ of positive tweets as negative in this sample. However, neutral and negative tweets were falsely characterized as positive at a rate of $34\%$ and $25\%$, respectively. Negative tweets are successfully classified at a rate of 75\%. Sarcasm remains very difficult to detect (partially explaining the $25\%$ false positive), but it typically appears in a minority of tweets.

\begin{figure}[ht]
\centering
  \includegraphics[width = \textwidth]{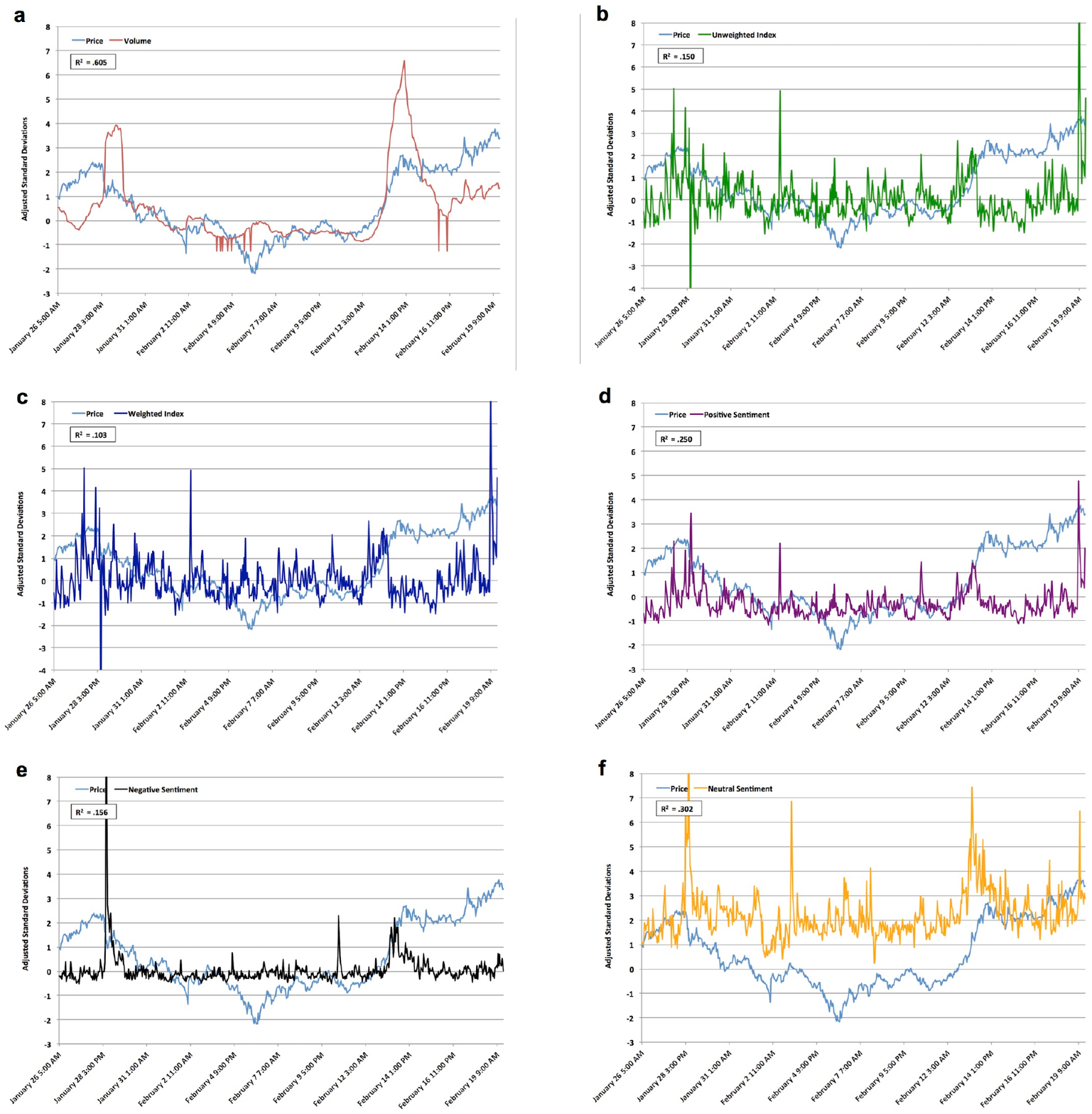}
  \caption{Shown are the price fluctuations versus our chosen six features, respectively, over the time period under consideration. (a) Price vs. Volume, (b) Price vs. Unweighted Index, (c) Price vs. Weighted Index, (d) Price vs. Pos. Sentiment, (e) Price vs. Neg. Sentiment, and (f) Price vs. Neutral Sentiment.}
  \label{pricefeature}
 \end{figure}

Having set the sentiment classification algorithms in place, we decided to train our model using six different features: Pure Positive Sentiment, Pure Negative Sentiment, Neutral Sentiment, An Unweighted Sentiment index, A Weighted Sentiment Index, and Hourly Trading Volume. These six features proved to be varied enough to train the model effectively on a variety of different trading points and resulted in the best and most accurate overall correlation with the testing data (as shown in Table~\ref{featurecorr}). The detailed co-plots of the different features versus the price curve over the study period is shown in Fig.~\ref{pricefeature}.

\begin{table}[ht]
\centering
\begin{center}
 \begin{tabular}{|c | c|} 
 \hline
     & Correlation Coefficients \\
 \hline
 Trading Volume & 0.605 \\ 
 \hline
 Neutral & 0.302 \\
 \hline
 Positive & 0.250 \\
 \hline
 Negative & 0.156 \\
 \hline
 Unweighted Index & 0.150 \\ 
 \hline
 Weighted Index & 0.103 \\
 \hline
\end{tabular}
\end{center}
\caption{5-factor correlation coefficients between the chosen feature and the price data, respectively.}\label{featurecorr}
\end{table}

\begin{figure}[ht]
\centering
  \includegraphics[width = 0.8\textwidth]{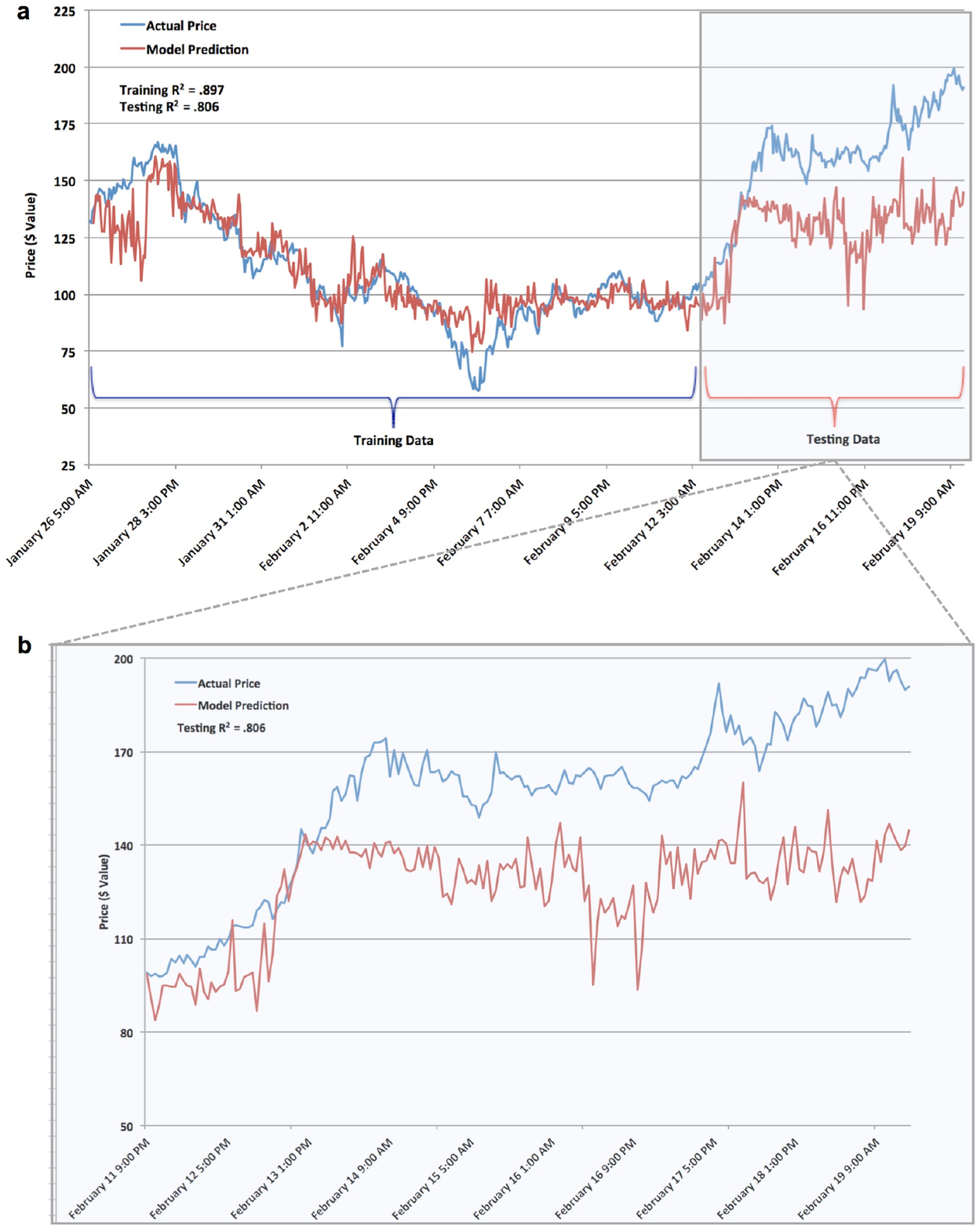}
  \caption{Comparison of model prediction and actual price data. Panel (a) plots the fitted price curve obtained from the training price data and the predicted price curve with respect to the testing data. Panel (b) details the model prediction price data as compared to the testing real price data.}
  \label{pricefit}
 \end{figure}

In testing our model, we were able to produce price data that strongly reflected the actual fluctuations (see Fig.~\ref{pricefit}). In particular, it is significant that our model achieved a Pearson correlation of .806 when tested against the actual test data, yielding a statistical significance at the $p \textless 0.0001$ level. As such, our model provides a viable method to predict price fluctuations, and also serves as a proof of concept that statistical analyses using Twitter sentiment can also be used to analyze price fluctuations in additional cryptocurrencies. It is also interesting to note that despite the similar directionality between the price model and actual price fluctuations, there appears to be a price \emph{€œgap}€ between the two of around \$30 (see Fig.~\ref{pricefit}b). One possible explanation to this gap is the discrepancies between the training and testing data (as summarized in Table~\ref{traintest}). First, it is important to note that the model was trained on data that primarily exhibited a negative trend (see Table~\ref{traintest}). As such, it is possible that the model became more desensitized to positive stimuli, and more sensitive to negative stimuli. In the testing data, however, the model was exposed to $\sim$ 3\% decrease in positive stimuli and $\sim$ 0.5\% increase in negative stimuli (Table~\ref{traintest}). The number of average tweets per hour also increased by $\sim$ 15\% (Table~\ref{traintest}). As such, it is possible that the model reacted to the change in these factors by exhibiting a slightly lower price expectation than what the actual market reflected. However, the overall directionality and correlation within the model remained strong, suggesting that if the model were also trained on data that exhibited positive trends, a more accurate set of predictions would have resulted.

\begin{table}[ht]
\centering
\begin{center}
 \begin{tabular}{|c |c c c c|} 
 \hline
     & Positive & Negative & Neutral & Total \\
 \hline
 Average Hourly Tweets in Training Data & 73.7 & 18.8 & 75.2 & 167.7\\
 \hline
 Percentage of Total & 43.9\% & 11.2\% & 44.8\% & -\\
 \hline
 Average Hourly Tweets in Testing Data & 82.4 & 23.5 & 95.5 & 201.4\\
 \hline
 Percentage of Total & 40.9\% & 11.7\% & 47.4\% & -\\
 \hline
\end{tabular}
\end{center}
\caption{Discrepancies of Twitter sentiments between testing and training data.}
\label{traintest}
\end{table}

\section{Discussion and conclusions}

In conclusion, our results suggest that by analyzing Twitter sentiment and trading volume, an Extreme Gradient Boosting Regression Tree Model serves as a viable means of predicting price fluctuations within the ZClassic cryptocurrency market. As such, given the complete lack of research within this academic sphere, our model serves as a proof of concept that social media platforms such as twitter can be used to capture investor sentiment, and that this sentiment is an early signal to future price fluctuations in alternative cryptocurrencies. Of particular interest is seeing whether this approach produces similarly strong results when applied to other alternative cryptocurrencies such as ZCash and Bitcoin Private. However, this discovery sheds light to the possibility of arbitrage opportunities that utilize social media platform sentiment to predict future cryptocurrency prices.

Our pricing model could be further improved by factoring in other social media platforms or data, such as Google Search results, Facebook posts, and Reddit Posts. Moreover, the dictionary that we have used in our model could be also be aided by adding investment-specific terms that indicate positive and negative sentiment such as ``bull'' and ``bear'' respectively. As seen from our manual vs. algorithm cross-analysis, the algorithm's greatest weakness is in classifying tweets that should otherwise be characterized as ``negative'' as ``positive.'' After careful review it is evident that such inaccurate characterizations are due to the algorithm's inability to detect sarcasm, a pervasive language schema in popular social media platforms. As such, further research to enhance our algorithm to detect sarcasm would increase the reliability of the sentiment analysis, and thereby potentially improve the accuracy of our prediction to retail driven price changes. 

Lastly, it would be interesting to further train and test our model over a longer time period. Given the confines of the date of our cryptocurrency's fork and our computational capacity, our study was restricted to a data set that covered a time frame of $3.5$ weeks. However, our results suggest a necessity to devote further resources and investments that would enable us to implement study our pricing model under a longer time frame and with other cryptocurrencies. 

\section*{Acknowledgments}
F.F. gratefully acknowledges support from the Dartmouth Faculty Startup Fund, Walter \& Constance Burke Research Initiation Award, and NIH Roybal Center Pilot Grant.

\end{document}